# Exfoliation of Two-Dimensional Nanosheets of Metal Diborides

Ahmed Yousaf,[1,2,†,‡] Matthew S. Gilliam,[1,2,‡] Shery L. Y. Chang,[3] Mathias Augustin,[4] Yuqi Guo,[5] Fraaz Tahir,[6] Meng Wang,[7] Alexandra Schwindt,[7] Ximo S. Chu,[5,†] Duo O. Li,[5] Suneet Kale,[1,2,5] Abhishek Debnath,[1,2] Yongming Liu,[6] Matthew D. Green,[7] Elton J. G. Santos,[4*] Alexander A. Green,[1,2*] and Qing Hua Wang[5*]

1. Biodesign Center for Molecular Design and Biomimetics, The Biodesign Institute, Arizona State University, Tempe, Arizona 85287, United States

2. School of Molecular Sciences, Arizona State University, Tempe, Arizona 85287, United States

3. Eyring Materials Center, Arizona State University, Tempe, Arizona 85287, United States

4. School of Mathematics and Physics, Queen's University Belfast, Belfast BT7 1NN, United Kingdom

5. Materials Science and Engineering, School for Engineering of Matter, Transport and Energy, Arizona State University, Tempe, Arizona 85287, United States

6. Aerospace Engineering and Mechanical Engineering, School for Engineering of Matter, Transport and Energy, Arizona State University, Tempe, Arizona 85287, United States

7. Chemical Engineering, School for Engineering of Matter, Transport and Energy, Arizona State University, Tempe, Arizona 85287, United States

‡These authors contributed equally.

*Corresponding authors: qhwang@asu.edu, alexgreen@asu.edu, e.santos@qub.ac.uk

†Present address: Intel Corp.

ABSTRACT. The metal diborides are a class of ceramic materials with crystal structures consisting of hexagonal sheets of boron atoms alternating with planes of metal atoms held together with mixed character ionic/covalent bonds. Many of the metal diborides are ultrahigh temperature ceramics like $HfB_2$, $TaB_2$, and $ZrB_2$, which have melting points above 3000ºC, high mechanical hardness and strength at high temperatures, and high chemical resistance, while $MgB_2$ is a superconductor with a transition temperature of 39 K. Here we demonstrate that this diverse family of non-van der Waals materials can be processed into stable dispersions of two-dimensional (2D) nanosheets using ultrasonication-assisted exfoliation. We generate 2D nanosheets of the metal diborides $AlB_2$, $CrB_2$, $HfB_2$, $MgB_2$, $NbB_2$, $TaB_2$, $TiB_2$, and $ZrB_2$, and use electron and scanning probe microscopies to characterize their structures, morphologies, and compositions. The exfoliated layers span up to micrometers in lateral dimension and reach thicknesses down to 2-3 nm, while retaining their hexagonal atomic structure and chemical composition. We exploit the convenient solution-phase dispersions of exfoliated $CrB_2$ nanosheets to incorporate them directly into polymer composites. In contrast to the hard and brittle bulk $CrB_2$, we find that $CrB_2$ nanocomposites remain very flexible and simultaneously provide increases in the elastic modulus and the ultimate tensile strength of the polymer. The successful liquid-phase production of 2D metal diborides enables their processing using scalable low-temperature solution-phase methods, extending their use to previously unexplored applications, and reveals a new family of non-van der Waals materials that can be efficiently exfoliated into 2D forms.

The metal diborides are a family of ceramic compounds with the general formula $MB_2$, where M can be a variety of metals from Groups II, IVB and VB such as Hf, Cr, Ti, Zr, Mg, etc. The most common crystal structure for these materials is the $AlB_2$ type, belonging to the *P6/mmm* space group symmetry, which has the boron atoms in covalently bound hexagonal sheets separated by layers of close-packed metal atoms,[1] as shown in **Figure 1a**. The interactions between the alternating layers of metal atoms and boron atoms have mixed ionic and covalent character. Depending on the identity of the metal atoms, the M-B and B-B bond lengths and bonding strengths vary, leading to dramatic composition-dependent differences in properties.[2,3] The metal diborides exhibit a diverse range of remarkable physical properties.[1,4–6] For example, several of them have very high melting points, including some that exceed 3000ºC, such as $HfB_2$ and $ZrB_2$ which are considered ultrahigh temperature ceramics (UHTCs).[1,7,8] Other examples of remarkable properties exhibited by the metal diborides include high mechanical strength and hardness (Young's modulus above 480 GPa for $ZrB_2$ and $HfB_2$), exceptional mechanical performance at temperatures above 1500ºC for $ZrB_2$ ceramics,[9] high thermal conductivity,[10] resistance to chemical attack,[1,11,12] high fracture toughness,[13] high wear resistance,[14] high electrical conductivity,[15] and electrocatalytic activity for the hydrogen evolution reaction.[16] In contrast, $MgB_2$ has a substantially lower melting temperature of 830°C but is well known as a superconductor with a transition temperature of 39 K, which ranks among the highest of the conventional superconductors.[17–19]

These properties make the metal diboride UHTC materials suitable for extreme conditions, with applications such as high temperature electrodes, coatings in cutting tools, neutron absorption in nuclear reactors, armor, and components for hypersonic flight and atmospheric reentry vehicles.[1,7,20] However, these ceramic materials have been mostly limited to uses as rigid structures. Furthermore, processing of the metal diborides usually require methods such as high temperature sintering and pressing,[21] sintering with various additives,[22,23] and spark plasma sintering,[24,25] but the densification of the UHTC metal diborides, which is required for aerospace and nuclear applications, is difficult due to their high melting temperatures.[21] Thus, new methods of processing the metal diborides have the potential to greatly expand the range of applications for these materials, enabling their exceptional properties to be fully exploited. In recent years, the metal diborides along with other boron-based compounds, have been gaining renewed attention from chemists who have been studying their electrochemical properties and potential for alternative low-temperature and solution-based processing routes.[26–34]

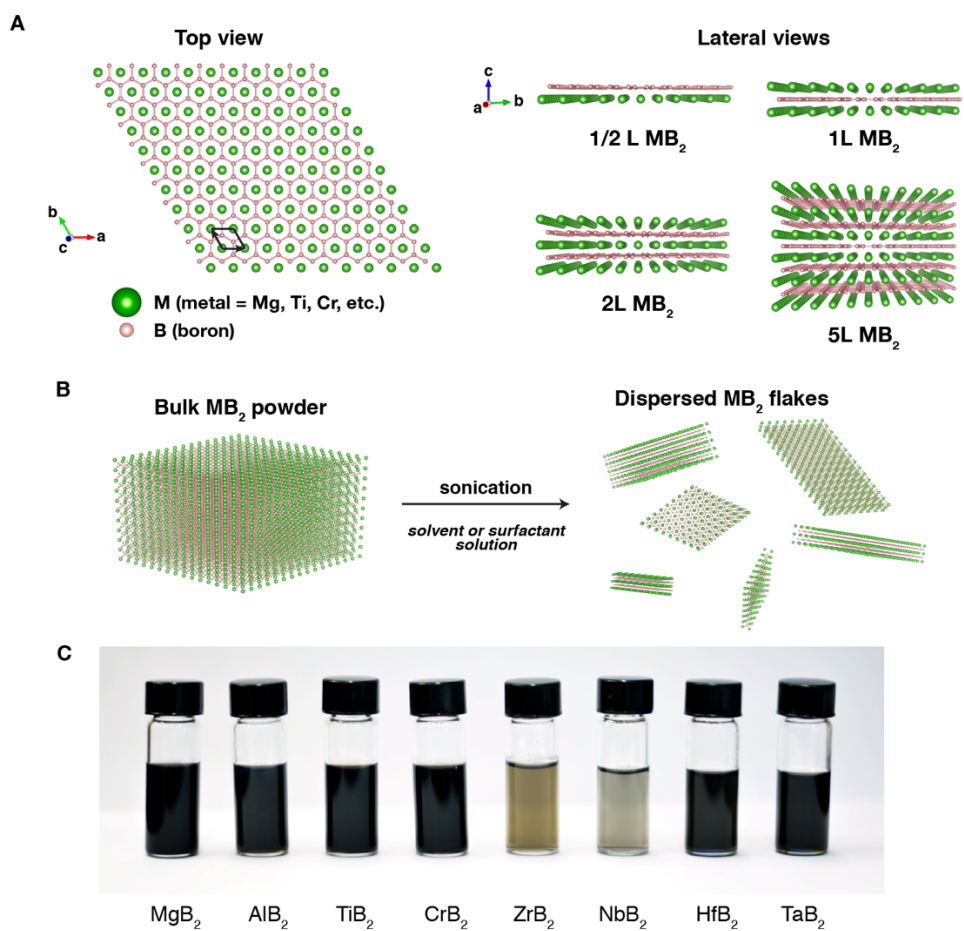

**Figure 1. Structure and dispersions of metal diborides.** (**A**) Structure of metal diborides in top view and lateral views of different layer thicknesses. One unit cell with **a** and **b** axes is outlined in the top view. (**B**) Schematic illustration of exfoliation process employed to disperse metal diborides in solvents or surfactant solutions via sonication. (**C**) Photograph of metal diboride dispersions in various organic solvents and aqueous surfactant solutions: $MgB_2$ in dimethylformamide (DMF), $AlB_2$ in DMF, $TiB_2$ in sodium cholate (SC) aqueous solution, $CrB_2$ in SC, $ZrB_2$ in myristyltrimethylammonium bromide (MTAB) aqueous solution, $NbB_2$ in MTAB solution, $HfB_2$ in N-methyl-2-pyrrolidone (NMP), and $TaB_2$ in NMP.

The layered structure of the metal diborides with their graphene-like boron sheets suggests the possibility of exfoliation into thin 2D nanosheets. The ability to produce metal diborides as 2D nanosheets using mild processing conditions could provide an attractive new route to applying these materials and extending their use to new contexts. Despite this potential, efforts to produce 2D metal diborides have so far been focused mainly on the superconducting $MgB_2$ rather than any of the UHTC materials, and yielded chemically modified products that have different structures, stoichiometries, and chemical functionalities than their bulk source materials. For example, Das et al. produced hydroxyl-functionalized and Mg-deficient few-layer B-based sheets by sonicating bulk $MgB_2$ powder in water.[31] The same group followed up with some other strategies for producing highly modified nanosheets including a chelation-assisted exfoliation utilizing ethylenediamine tetraacetic acid (EDTA), which removes much of the Mg metal and adds hydride, hydroxyl and oxy-functional groups to the remaining B-based layers;[34] an ionic liquid-assisted exfoliation in 1-butyl-3-methyl-imidazolium tetrafluoroborate;[33] and treatment with sulfuric acid followed by intercalation with organoammonium ions.[32] Nishino et al. similarly showed that sonication of $MgB_2$ in water produces hydroxyl-functionalized Mg-deficient boron nanosheets,[35] and that reaction of $MgB_2$ with an ion exchange resin in an aqueous solution produces hydrogen boride (HB).[36] For the non-$MgB_2$ metal diborides, Lim et al. treated $TiB_2$ with butyllithium and sodium naphthalinide but were unsuccessful in its exfoliation.[27] Thus, to our knowledge, a successful method to produce unmodified few-layer 2D forms of the family of metal diborides has not been reported.

Here, we report the efficient exfoliation of the family of metal diboride compounds into unmodified few-layer 2D nanosheets using direct liquid-phase exfoliation. Using ultrasonication in suitable solvents or aqueous surfactant solutions, we successfully exfoliate eight different metal diboride compounds – $AlB_2$, $CrB_2$, $HfB_2$, $MgB_2$, $NbB_2$, $TaB_2$, $TiB_2$, and $ZrB_2$ – to prepare liquid-phase dispersions with concentrations up to 2.4 mg mL$^{-1}$. We use density functional theory (DFT) calculations to demonstrate that the metal diborides $MgB_2$ and $HfB_2$ are stable in few-layer-thick 2D forms and to simulate their electron energy loss spectra (EELS). Transmission electron microscopy (TEM) imaging of the resulting dispersions confirm the 2D nature of the nanosheets and reveal that their hexagonal structure is retained after exfoliation. Atomic force microscopy (AFM) of the nanosheets show their nanometer-scale thickness. Furthermore, EELS obtained from the 2D metal diboride nanosheets demonstrate that the chemical composition is maintained after the exfoliation process. Lastly, we employ our direct exfoliation approach to prepare dispersions of 2D chromium diboride ($CrB_2$) directly in aqueous solutions of polyvinyl alcohol (PVA), which enables facile production of metal diboride reinforced polymer composites. The resulting $CrB_2$/PVA composites provide improvements in elastic modulus and ultimate tensile strength of up to 23% and 48%, respectively, over the polymer alone. These improvements in mechanical properties exceed that achieved by the direct incorporation of any 2D materials so far, such as graphene and $MoS_2$, to PVA, even after size-sorting to select for the largest nanosheets.[37–39] These results introduce a new family of normally brittle metal diboride ceramics that are amenable to scalable, low-temperature solution-phase processing, thus enabling their use in flexible, conformal, and stretchable forms.

## Results

### *Exfoliation of $MB_2$ nanosheets*

Few-layer metal diboride nanosheets were prepared using ultrasonication in several solvents and aqueous surfactant solutions, as schematically illustrated in **Figure 1B**. Ultrasonication relies on the principle of cavitation to shear apart the sheets, which are then stabilized by the surrounding solvent or surfactant molecules. Bulk powders of each of the metal diborides were added to solvents or surfactant solutions, and sonicated by a tip or bath ultrasonicator (see further details in the *Methods* section). Following centrifugation to remove poorly dispersed materials, the supernatant was decanted. This process was applied to eight different metal diborides for which powder source materials were available: aluminum diboride ($AlB_2$), chromium diboride ($CrB_2$), hafnium diboride ($HfB_2$), magnesium diboride ($MgB_2$), niobium diboride ($NbB_2$), tantalum diboride ($TaB_2$), titanium diboride ($TiB_2$), and zirconium diboride ($ZrB_2$).

The resulting solution-phase dispersions were grey to dark black and remained stable in suspension for weeks without precipitating with the exception of $AlB_2$, which precipitated after 2-3 days. **Figure 1C** shows a photograph of dispersions of the different metal diborides prepared using this method. The optical extinction spectra, which are a combination of absorption and scattering, obtained from the dispersions are mostly featureless with the exception of $MgB_2$, which shows two broad peaks near 400 nm and 850 nm as shown in the Supplementary Information, **Figure S1a**. Dimethylformamide (DMF) was found to be an effective solvent for $MgB_2$ and $AlB_2$, while N-methyl-2-pyrolidone (NMP) was effective for $HfB_2$ and $TaB_2$. $TiB_2$ and $CrB_2$ were efficiently dispersed in aqueous solution using the anionic surfactant sodium cholate (SC) and $ZrB_2$ and $NbB_2$ were best exfoliated in aqueous solution using the cationic surfactant myristyltrimethylammonium bromide (MTAB). The concentrations of the metal diboride liquid-phase dispersions were determined using inductively coupled plasma mass spectrometry (ICP-MS). These measurements showed a broad range of concentrations from 2.4 mg/mL for $MgB_2$ to 0.07 mg/mL for $ZrB_2$ (Supplementary Information, **Figure S1b**). These values are comparable to the concentrations reported for previous studies of the most widely studied 2D materials based on vdW solids, such as graphene, BN, and transition metal dichalcogenides.[40–43]

### *Structural characterization of MB$_2$ nanosheets*

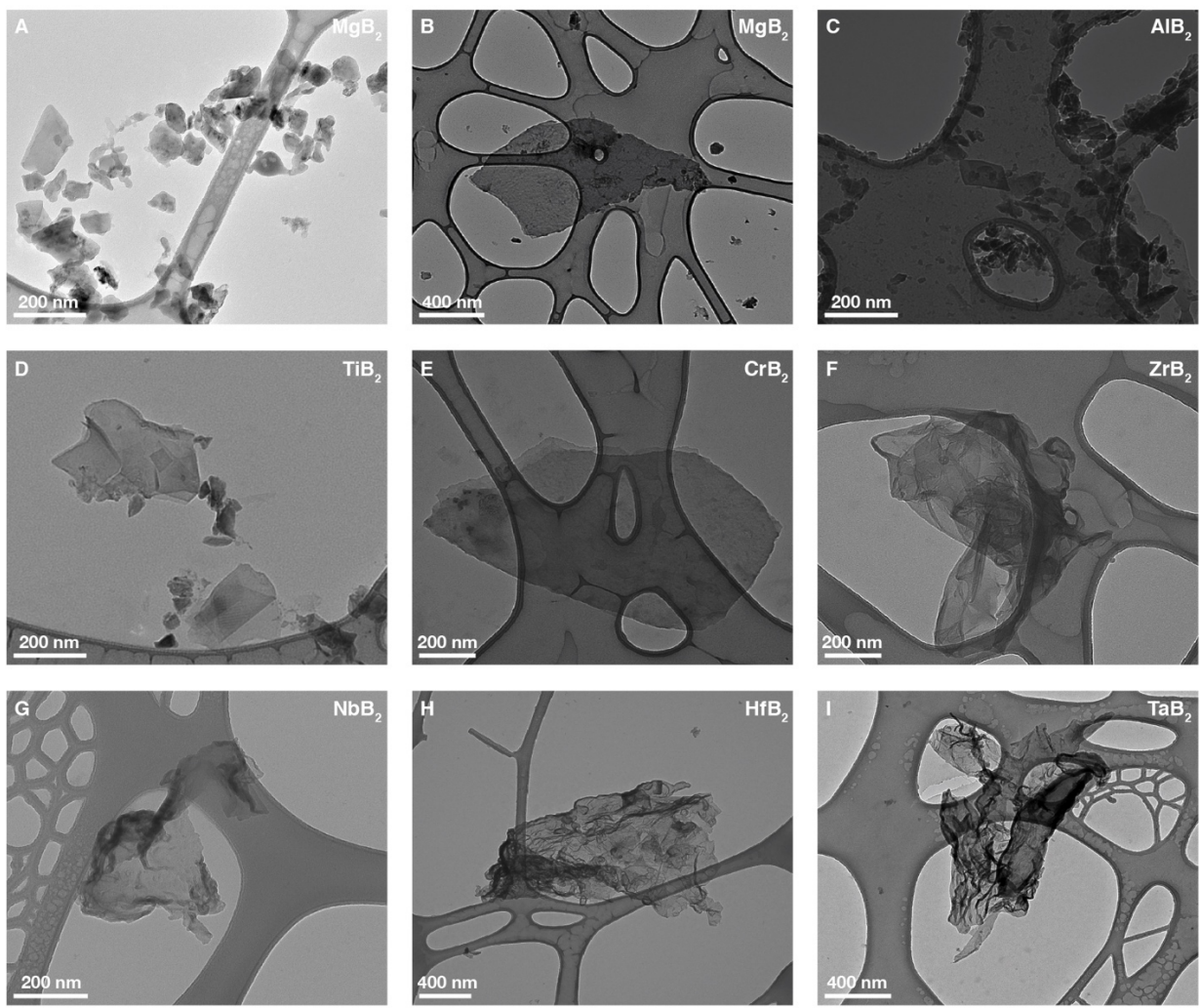

**Fig. 2. TEM imaging of metal diboride flakes.** (**A**)-(**I**) Low magnification TEM images of flakes from eight different metal diboride compounds generated by solution phase dispersions. Lateral dimensions of flakes range from 30 nm to 2 $\mu$m.

To confirm the 2D nature of the sheets, transmission electron microscope (TEM) images were obtained for all eight metal diborides (**Figure 2**). The flakes are found in different morphologies varying from flat, planar sheets, as shown for the $MgB_2$, $AlB_2$, $TiB_2$, and $CrB_2$, to folded and crumpled ones, as shown for $ZrB_2$, $NbB_2$, $HfB_2$, and $TaB_2$. The flake sizes were found to vary, with $AlB_2$ having the smallest flakes (< 100 nm across) and $MgB_2$ the largest up to several microns across. Additional TEM images are shown in the Supplementary Information, **Figure S2**. Histograms of the areas of $HfB_2$ and $MgB_2$ flakes from TEM images were plotted and shown in the Supplementary Information, **Figure S4**. We observe a decrease in the average flake area with longer centrifugation times.

Atomic force microscopy (AFM) was used to characterize the thicknesses and areas of nanosheets of $MgB_2$, $TiB_2$, $HfB_2$, and $ZrB_2$ (**Figure 3**). Samples of metal diboride nanosheets were prepared by both tip sonication and bath sonication, and deposited by spin coating onto $SiO_2$/Si and mica substrates (see Methods section for more details). We found that this process in general resulted in flakes of smaller lateral area to be deposited onto the substrate compared to those observed in TEM imaging, but that larger areas could be obtained from a less aggressive bath sonication treatment to reduce the likelihood of sonication-induced cutting. AFM images in **Figure 3** show flakes of different sizes and varying thicknesses. Height profiles show typical thicknesses ranging from about 3 to 18 nm and lateral dimensions of about 150 nm to 4 μm across. Given the interlayer spacing of 0.35 nm for bulk crystalline $HfB_2$, $ZrB_2$, and $MgB_2$,[28] and assuming a hydration layer of 0.35 to 1.5 nm thickness on the outer surface of the flakes,[44–46] these nanosheets are about 8-50 layers thick. Additional AFM images of the sample areas surrounding these flakes with more flakes of various sizes are shown in the Supplementary Information, **Figure S3**. Histograms of the areas and thicknesses of $HfB_2$ and $MgB_2$ flakes were plotted and shown in the Supplementary Information, **Figure S5**. Generally, thicker flakes have larger areas, and we achieve much thinner flakes for $MgB_2$ than $HfB_2$, achieving a significant proportion of flakes thinner than 5 nm.

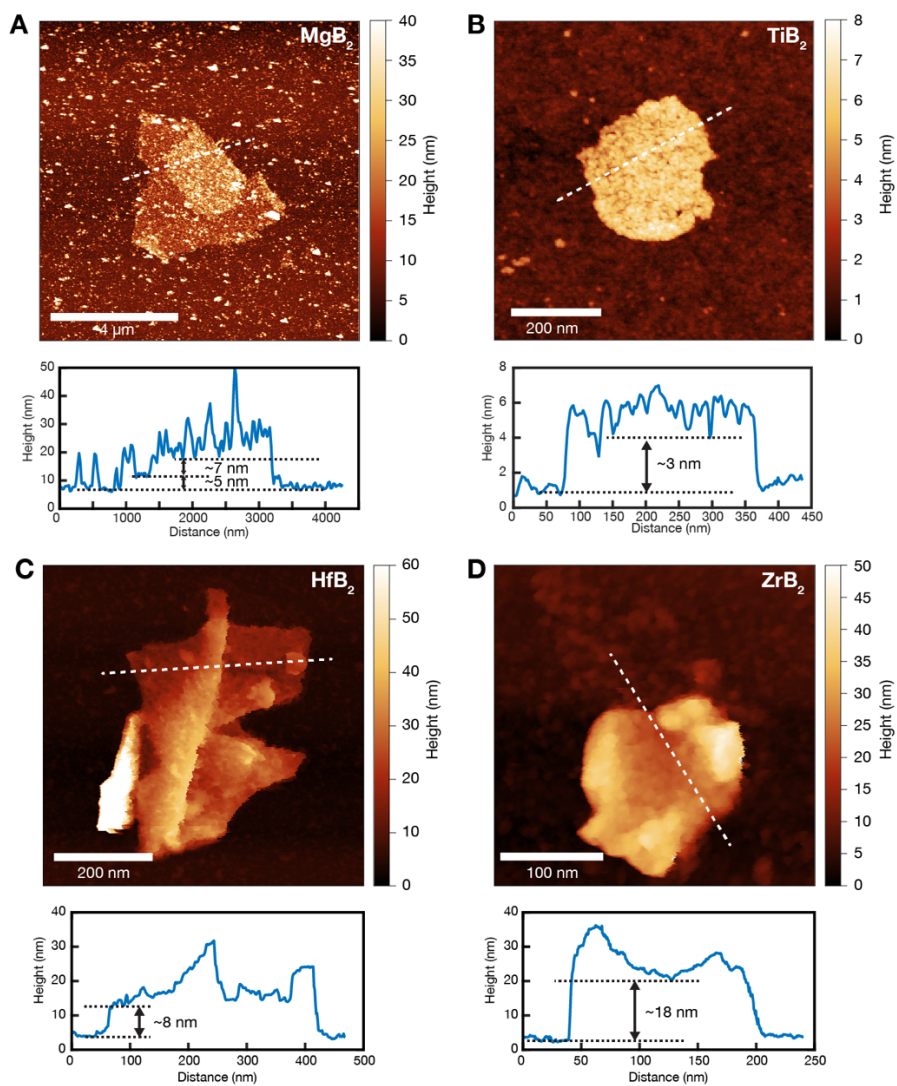

**Figure 3. AFM imaging of metal diboride flakes.** (**A**) $MgB_2$ flake dispersed by bath sonication in IPA, deposited onto $SiO_2/Si$ substrate and imaged in air. (**B**) $TiB_2$ flake dispersed by bath sonication in IPA, deposited onto $SiO_2/Si$ substrate and imaged in air. (**C**) $HfB_2$ flake dispersed by tip sonication in isopropanol, deposited on mica substrate and imaged in air. (**D**) $ZrB_2$ flake dispersed by tip sonication in isopropanol, deposited on $SiO_2/Si$ substrate and imaged in water. Height profiles along the dashed lines are shown below each AFM image.

X-ray diffraction (XRD) was also conducted according to the details described in the Methods section for the exfoliated $MgB_2$, $ZrB_2$, and $CrB_2$ and compared to their corresponding initial bulk powders, as shown in the Supplementary Information, **Figure S8**. The main diffraction peaks are almost entirely absent in the exfoliated materials, indicating that they have lost their stacking order.

In order to elucidate the atomic structure of the metal diboride nanosheets, we carried out aberration-corrected high-resolution TEM (ACTEM) studies of $HfB_2$. **Figure 4A** shows a multilayer $HfB_2$ flake composed of 4-5 layers. There are some smaller domains of different periodicities, which we attribute to smaller flakes of $HfB_2$ restacking onto the larger flake at different crystal alignments. Imaging of the same sample at increased magnification in the region marked by a square (**Figure 4B**) revealed the hexagonal structure of the $HfB_2$, similar to that of the bulk crystal form. For more detailed comparison with bulk $HfB_2$, we conducted electron scattering simulation using the multislice method[47,48] to simulate the expected HRTEM image of $HfB_2$, with the simulation parameters corresponding to the experimental imaging conditions. We found that the experimental images closely match the simulated structure viewed along the [001] direction (**Figure 4B and 4C**). Additional HRTEM images and EELS spectra of $MgB_2$ are shown in the Supporting Information (**Figure S7**). Calculated geometrical parameters using density functional theory (DFT) simulations including vdW interactions (see *Methods* section for details) of $MgB_2$ and $HfB_2$ layers of different thicknesses are shown in **Table S2** in the Supplementary Information. The experimental HRTEM images showed {100} lattice spacing of $MgB_2$ and $HfB_2$ spacings of 2.68 ± 0.05 Å and 2.72 ± 0.05 Å, respectively. After converting the lattice spacings to the Mg-Mg and Hf-Hf bond length given that the metal atoms are arranged in planes, we find that the experimental measurements are in good agreement with the DFT calculated values (see **Supporting Information**, **Table S2**). It is also worth mentioning that the phonon dispersion for the exfoliated layers does not show any imaginary or soft modes that could compromise the integrity of the structure (**Figure S10**). This indicates a robust stability for the nanosheets.

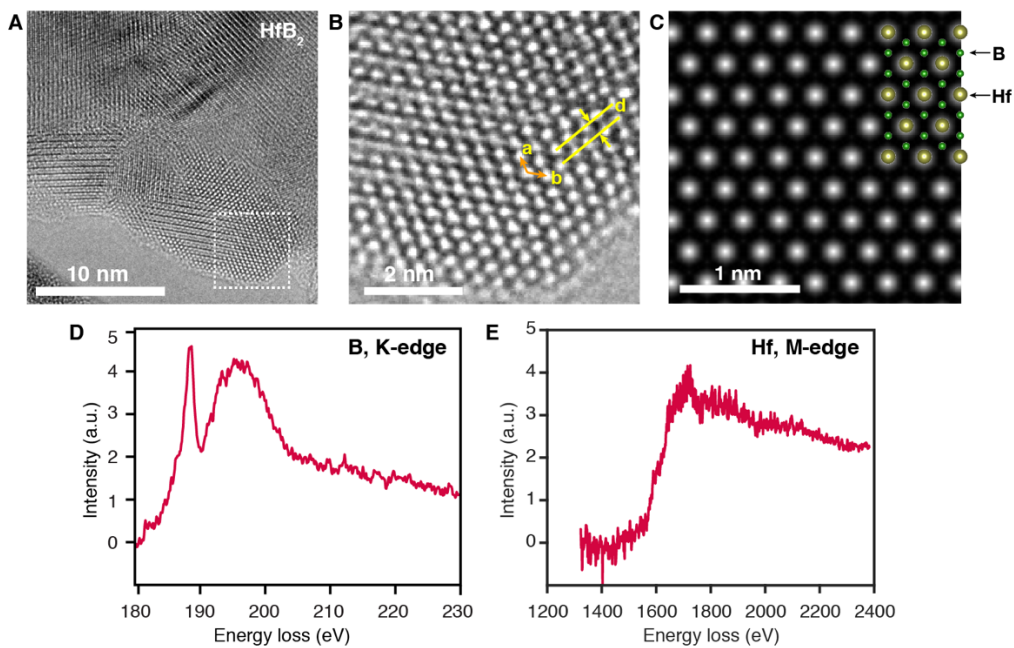

**Figure 4. HRTEM and EELS characterization of HfB₂ flakes.** (**A**) Aberration corrected TEM imaging of HfB₂ of 4-5 layers. (**B**) Enlarged image of the area marked by a square in (A), with the arrows depicting the basis vectors of the hexagonal structure of HfB₂. The parallel lines indicate the (100) planes, which have a spacing of 2.68 Å. (**C**) Simulated HRTEM image of 4 layers of HfB₂ based on the bulk crystal structure, with HfB₂ structure overlaid. (**D**)-(**E**) EELS spectra of HfB₂, with the B K-edge in (D) and the Hf M-edge in (E).

### Compositional analysis of MB₂ nanosheets

Electron energy loss spectroscopy (EELS) was used to identify the elemental makeup of the solution-dispersed metal diborides. The B K-edge for HfB₂ in **Figure 4d** shows the characteristic peaks of diboride. The peak at 188 eV corresponds to the transitions to the π* antibonding state, which originates from the $sp^2$ bonding present within the boron layers. The broad peak centered at 196 eV can be associated with the transitions to the σ* antibonding orbitals.[49] In **Figure 4e** the broad peak above 1600 eV corresponds to the Hf M-edge confirming the presence of Hf.[50] Furthermore, quantitative composition analysis based on the spectra containing B and Hf edges shows the Hf:B ratio to be about 1:2.09, which is very close to the expected value of 1:2 (see Supplementary Information, **Figure S6**). These results show that the chemical composition of pristine HfB₂ is generally maintained after solution-phase exfoliation, although there is a small degree of oxidation, which is also shown in the energy-dispersive X-ray spectroscopy (EDS) spectrum in **Figure S6**. These spectra were taken at multiple locations on the nanosheet and little to no spatial variations were found. We also carried out similar atomic structural and compositional analyses of $MgB_2$ sheets as described in the Supporting Information, **Figure S7**. Calculated EELS spectra for bulk and monolayer $MgB_2$ (**Figure S11** and **Figure S12**) have features that agree qualitatively with the experimental results.

### *Synthesis and mechanical testing of CrB₂/PVA composites*

While structures based on metal diboride compounds are conventionally processed at very high temperatures,[1,51] we use liquid-phase exfoliation to produce dispersions that can be incorporated into polymer composites at room temperature. Furthermore, it produces few-layer-thick metal diboride nanosheets that are likely substantially less brittle than their bulk counterparts. As a demonstration of the utility of this scalable sample preparation method, we directly integrated the exfoliated 2D nanoflakes into polymer composites using the commonly used and water-soluble matrix poly(vinyl alcohol) (PVA). Two-dimensional materials have been extensively studied as filler materials for mechanical reinforcement of polymers, with PVA as a commonly chosen polymer due to its ease of processing and compatibility with liquid phase exfoliated dispersions. Graphene, in particular, has attracted considerable interest as a result of its remarkable strength and stiffness.[52] The exceptional mechanical properties of the bulk metal diborides suggest their use as filler materials, provided that they can be processed into suitable forms. We thus screened the metal diborides to determine which could produce stable dispersions through direction ultrasonication in 1% w/v aqueous PVA. We found that $CrB_2$ produced the most stable and highest concentration dispersions. $CrB_2$ also has good wear resistance,[14] high hardness, high melting point (2200°C), and high performance under extreme temperature and pressure conditions.[53] These $CrB_2$/PVA dispersions were concentrated and formed into films by a casting process (see **Methods** for more details). The dilution ratio was varied to obtain composites with different mass fractions of $CrB_2$. XRD spectra of the $CrB_2$/PVA composite films have only peaks from PVA and none from the bulk $CrB_2$ phase, suggesting good incorporation of the $CrB_2$ nanosheets, as shown in the Supplementary Information (**Figure S8**). The mechanical properties of multiple composite membranes were measured by tensile testing (see **Methods** for more details).

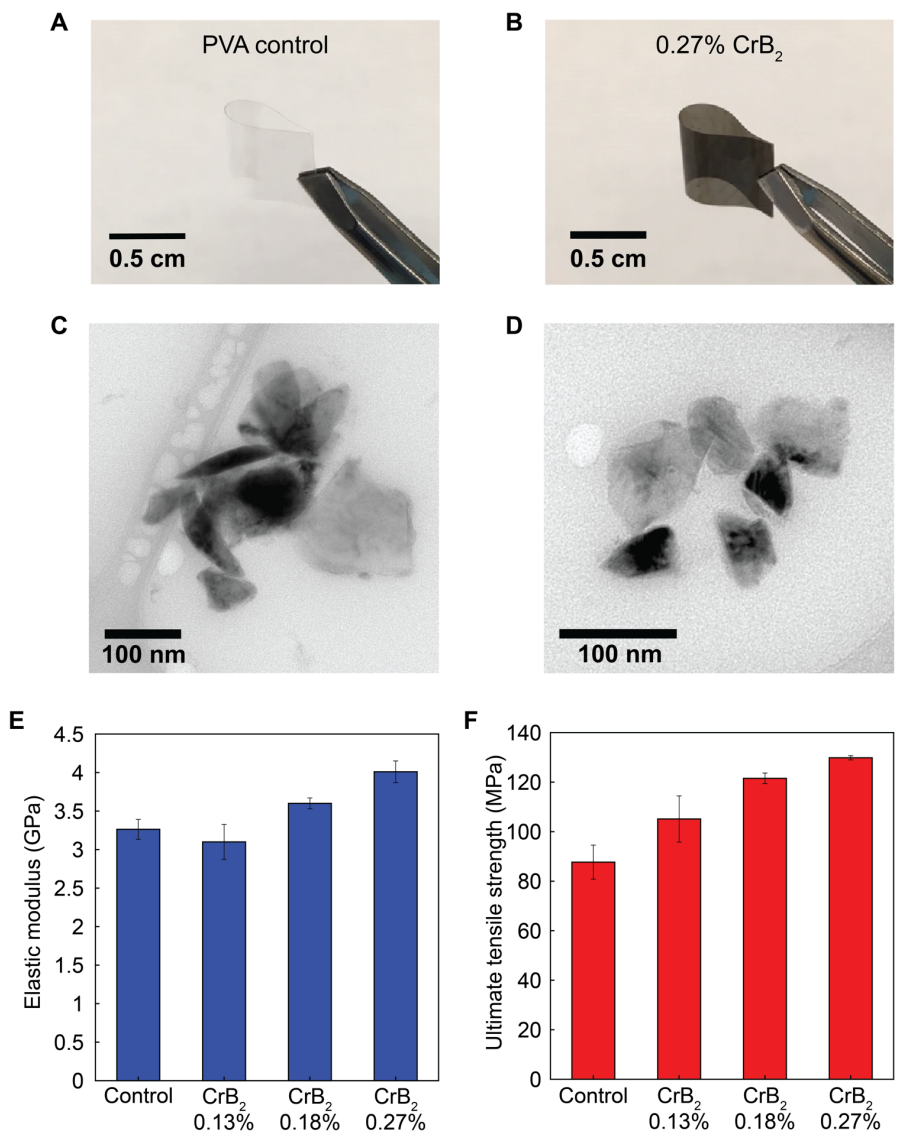

**Figure 5.** Polymer composites and mechanical properties. (**A**)-(**B**) Photographs of flexible PVA film (A) and PVA film containing ~0.27% $CrB_2$ (B). (**C**)-(**D**) Representative TEM images of $CrB_2$ nanosheets dispersed in PVA. (**E**) Elastic modulus for PVA control with no $CrB_2$, and with $CrB_2$ of different loadings. (**F**) Ultimate tensile strength values for PVA control with no $CrB_2$, and with $CrB_2$ of different loadings.

The polymer composites remained quite flexible with the addition of $CrB_2$ from 0.13% to 0.27% w/w loadings, as shown in the photograph **Figure 5B**. Representative stress-strain curves for PVA alone and PVA $CrB_2$ nanosheets are shown in the Supporting Information, **Figure S9**.

After measuring the mechanical properties from multiple samples, we found that the $CrB_2$/PVA composites with up to 0.27% of $CrB_2$ showed increases in the elastic modulus and ultimate tensile strength of up to 23% and 48%, respectively. Comparisons of the mechanical properties of the $CrB_2$/PVA composites to results from the literature for PVA composites reinforced by direct incorporation of other 2D materials are shown in the Supporting Information **Table S1**; our results are similar to the enhancement of mechanical properties seen for conventional 2D materials and carbon nanotubes. [37][54][55]

**Discussion**

We have demonstrated for the first time the solution-phase dispersion of chemically unmodified 2D nanosheets of eight different metal diborides using ultrasonication-assisted exfoliation of bulk powders. The dispersions of metal diborides have concentrations as high as 2.4 mg mL$^{-1}$. We have found that some solvents are more effective at dispersing certain metal diborides than others, and we hypothesize that materials with different metal compositions have different resulting surface energies along the exposed cleavage planes, and that some planes are better stabilized by some solvents or surfactants than others. A more detailed investigation of different solvents for dispersing metal diborides is being undertaken. TEM and AFM imaging were used to confirm the 2D nature of the resulting nanomaterials, showing thin sheet morphologies with lateral dimensions ranging from 100 nm to multiple microns and thicknesses down to 2-3 nm, although there is a distribution of dimensions. Histograms of the TEM and AFM images show this distribution of flake sizes and thicknesses for different metal diboride compositions, which may indicate differences in metal–boron bond strengths. There are also shifts in the distribution of flake areas as a function of centrifugation time, suggesting possible routes for future size sorting of the flakes,[56–58] which has been shown to improve mechanical properties for polymer nanocomposites of other 2D materials.[37,59] HRTEM and EELS were used to show that the crystal lattice and chemical composition are preserved after exfoliation. DFT calculations of phonon dispersions of different layer thicknesses show stability for atomically thin layers of $MgB_2$. Harnessing the enhanced processability and increased flexibility of the metal diboride nanosheets, we prepared polymer composites with as-exfoliated $CrB_2$, which resulted in increases in the mechanical strength of the PVA. Moving forward, there are many previously reported strategies for polymer nanocomposites that incorporate carbon nanotubes and graphene oxide such as *in situ* chemical cross-linking and alignment of the nanoparticle inclusions that can be exploited to achieve even higher enhancements in mechanical properties.[60–62]

These results represent an important step in the solution-phase, low-temperature processing of ultrahigh temperature ceramics and open the door to taking advantage of the entire class of metal diboride materials in previously unexplored contexts. The bonding character of metal diborides is different from the layered materials that are related to most conventional 2D materials, which generally possess weak vdW forces between layers. In the metal diborides, the boron atoms serve as electron acceptors and the metal atoms behave as electron donors; this donor-acceptor interaction introduces the ionic character in the M-B bond. However, partial interaction of *d*-electrons and the formation of *spd*-hybrid configurations impart covalent character to the M-B bond leading to a complex bonding environment.[1] The ability to exfoliate these compounds by breaking their mixed character bonds represents a key advancement in our ability to prepare 2D materials from compounds possessing strong interlayer bonding rather than weak van der Waals bonding. The production of 2D nanosheets from exfoliation of non-van der Waals precursor solids has recently been demonstrated for the natural ores hematite ($Fe_2O_3$)[63] and ilmenite ($FeTiO_3$),[64] and represent a relatively unexplored area for identifying novel low-dimensional materials. These 2D nanosheets of eight different metal diborides exfoliated into stable, homogeneous and high-concentration dispersions suggest that this method can be generalized to produce two-dimensional diborides from other metal diboride precursors of similar crystal structure and superhard mechanical strength,[65–68] providing a whole new family of 2D materials with a distinct set of properties. Our approach to achieve these boron-based 2D materials also differs from epitaxially grown borophene layers on metal substrates, which rely on the underlying support and are not freestanding.[69–71]

The direct integration of $CrB_2$ nanosheets in polymer composites and the resulting mechanical reinforcement highlights the potential for using metal diborides in solution-phase processing, which can take advantage of the wide range of existing methods previously developed for other 2D nanomaterials[57,72,73,73–76] to enable their use in other formats.[77–81] Future work may include investigation of the exfoliation behavior and unique properties of other types of boride compounds,[82,83] as well as the use of size sorting techniques to further improve the performance by isolating larger flakes.[37,59]

## Methods

### *Materials*

All chemicals were used as received without further purification. $MgB_2$, $AlB_2$, $TiB_2$, $NbB_2$, $TaB_2$, polyvinyl alcohol (PVA) with molecular weight of 89,000-98,000 g/mol, sodium cholate (SC), myristyltrimethylammonium bromide (MTAB), *N,N*-dimethylformamide (DMF) and 1-methyl-2-pyrrolidinone (NMP) were purchased from Sigma Aldrich, $CrB_2$ from Alfa Aesar, and $ZrB_2$ and $HfB_2$ from Smart Elements. More details on the materials and chemicals are listed in the Supplementary Information.

### *Preparation of metal diboride dispersions*

Tip sonication method: 0.4 g of each metal diboride powder was added to a 15 mL centrifuge tube along with 6 mL of the organic solvent or aqueous surfactant solution (1% w/v). Then the mixture was tip sonicated (Branson Digital Sonifier 450D, 3 mm diameter tip) for 1 hour at 20% amplitude corresponding to 11-13 W of power output. The resulting dispersion was transferred into 1.75 mL tubes and centrifuged at 5000 rcf for 4 minutes and the top 1 mL of the dispersion was collected from each tube. The concentrations of each dispersion were measured by inductively coupled mass spectrometry (ICP-MS).

Bath sonication method: 0.4 g of each metal diboride powder (e.g. $TiB_2$) (but 0.2 g for $MgB_2$ because of its higher dispersion efficiency) was mixed in 6 mL isopropanol (IPA) in centrifuge tubes and bath sonicated at full power for 8 hours in iced water (Branson CPX2800H). The dispersion was centrifuged at 5000 rcf (or 10,000 rcf for $MgB_2$) for 150 s, then the top 4 mL of the dispersion was extracted.

### *Inductively coupled mass spectrometry (ICP-MS)*

For each dispersion, a 50-µL aliquot was acidified overnight in 2 mL of concentrated nitric acid following by a dilution of the matrix to 2 wt% of $HNO_3$ aqueous solution. The concentration was determined by inductively coupled plasma mass spectroscopy (ICP-MS) (iCap Q quadrupole, Thermo Fisher).

### *Optical extinction spectroscopy*

The dispersions were diluted as needed and the corresponding solvent or aqueous surfactant solution at the same concentration was used as the reference. The UV-Vis-NIR spectra were collected from 400 to 1000 nm in quartz cuvettes using a Jasco V-670 spectrophotometer.

### *Transmission electron microscopy*

TEM images in **Figure 2** and **Figure S2** were acquired on Philips CM 12 at 80 kV and aberration corrected HR-TEM images in **Figure 3** and **Figure S3** were obtained on FEI aberration-corrected Titan at 300 kV. The images were taken at negative $C_s$ imaging condition, where the spherical aberration is tuned to about -14 μm. To determine lattice spacings, the peak position for each atom within the flakes were fitted first, then least squares criteria were used to refine the lattice spacings of all equivalent {100} planes in the flake. This over-redundant measurement method improves the accuracy of the measurement for small flakes. Flakes with well-defined edges and that were easily distinguished were identified, and their areas measured in ImageJ. The histograms of their areas are shown in **Figure S4**.

### *EDX and EELS*

EDX spectra were also collected on the Titan. EELS was performed on ARM 200F equipped with an Enfinium spectrometer operated at 200 kV and also 80 kV. The lower voltage was used to avoid electron beam damage to the thin boride layers. The B K-edge was taken using a dispersion of 0.1 eV/ch whereas the Hg M-edge and Mg M-edge were 1 eV/ch.

### *Atomic force microscopy*

AFM imaging was carried out on a Multimode V system (Bruker Corp.) with ScanAsyst-Air tips (Bruker) in ScanAsyst noncontact mode. Gwyddion[84] was used for image processing. Dispersions of $HfB_2$, $ZrB_2$, and $TiB_2$ were prepared in isopropanol using the tip or bath sonication methods as described above. Substrates were either freshly cleaved mica, or $SiO_2/Si$ wafers cleaned by sonication in acetone then isopropanol for 5 min each, and blown dry by ultrapure nitrogen. A drop of each dispersion was spin coated on each substrate at 2500-3000 rpm for 1 min. Spin coating was repeated two to five times to increase the areal concentration of nanosheets. Prior to AFM imaging, the samples were annealed at 300˚C for 9 h to remove organic residues. Processed images were then used to identify flakes, measure their height profiles using Gywddion to obtain thicknesses, and measure their areas using ImageJ. Histograms of the thickness and area of flakes were then plotted.

### *Polymer composite preparation*

To synthesize composites with PVA, 1.3 g of chromium diboride was bath sonicated in 20 mL of 1% w/v aqueous PVA solution for 4-10 hours and the resulting suspension was distributed equally in 1.75 mL tubes and centrifuged at 5000 rcf for 5 minutes. Control solutions of just PVA were processed using identical conditions. To further concentrate the $CrB_2$ in PVA, the $CrB_2$ dispersions were centrifuged to pellet out the $CrB_2$ nanosheets, half of the supernatant was removed, and the pelleted nanosheets were redispersed in the remaining PVA solution with shaking and bath sonication. The concentration of the resulting dispersion was determined by ICP-MS and used along with UV-Vis-NIR measurements to calculate and extinction coefficient for $CrB_2$ dispersed in 1% PVA. The mass loadings of the composites were then determined using the calculated extinction coefficient and UV-Vis measurements. The $CrB_2$/PVA dispersions were mixed with appropriate amounts of 1% w/v aqueous PVA solution containing no $CrB_2$ by vortexing to obtain the required $CrB_2$ mass loadings and then the mixtures were bath sonicated for 20 minutes. Afterwards, 24 mL of the resulting mixtures were poured into petri dishes and dried in an oven at 60ºC for 48 hours under vacuum. The dried $CrB_2$/PVA membranes were peeled from the petri dishes, cut into rectangular pieces of roughly 5 cm x 1 cm and taped at the edges such that the working length was roughly 3 cm. The average thickness of each film was calculated from three measurements along the length of each film and used for tensile stress calculations. These films were then subjected to tensile testing. The same procedure was also used PVA blank controls without adding any nanomaterials.

### *Tensile testing*

The mechanical behavior of the $CrB_2$/PVA composites was characterized by tensile testing of the 3 cm × 1 cm rectangular strips prepared as described above. The tests were conducted on an Instron E3000 Test Instrument with a 1 kN load cell and self-aligning grips. The stress was calculated from the load cell reading and the initial cross-section area of each specimen. The strain was calculated from crosshead displacement. All test specimens were stretched at a constant crosshead speed of 0.5 mm/s. At least three specimens were tested for each nanosheet concentration. Since the stress-strain curves do not have a clearly linear region, the elastic modulus was calculated by applying a linear least-squares fit to different small regions of the stress-strain curve and finding the maximum slope for each sample.

### X-ray diffraction (XRD)

Powders of exfoliated $MgB_2$ were prepared by bath sonicating batches of 1.3 g amounts of $MgB_2$ in 20 mL aliquots of IPA for approximately 2 hours. The resultant mixtures were then distributed into 1.75-mL microcentrifuge tubes and centrifuged for 10 minutes at 5000 rcf to remove large aggregates, isolating the nanosheets within the supernatant. The supernatant containing nanosheets was then collected and subjected to additional centrifugation such that the nanosheets within the dispersion pelleted out almost completely. The supernatant of the spun down nanosheets was removed and 200 μL was added to the pelleted nanosheets to redisperse them. This process was repeated many times to concentrate the nanosheets within one solution. The concentrated solution was finally spun down, the supernatant was removed, and the pellet was left to dry overnight in an oven to obtain the powder used for powder x-ray diffraction (PXRD). The $ZrB_2$ powder was prepared similarly, except that instead of bath sonication, tip sonication was used (2 hours, 35% amplitude) on batches of 1.3 g of $ZrB_2$ in 22 mL of IPA. Bulk powders of $MgB_2$ and $ZrB_2$ were used as received. For the PXRD measurements, films of the powders were prepared on quartz substrates by placing powders onto the middle of the substrate and spraying IPA on both sides to allow the powders to soak up the solvent and spread across the substrate. After drying, PXRD spectra of the films were collected on a Siemens D5000 Powder X-Ray Diffractometer at a scan rate of 3 degrees per minute using copper K-α radiation.

Nanocomposite PVA films containing different loadings of $CrB_2$ nanosheets, along with a film of PVA alone, were prepared as described earlier for mechanical testing samples. The films were then taped by the edges to a quartz substrate to collect PXRD spectra. Bulk $CrB_2$ powder was analyzed using by using a film prepared as above for the $MgB_2$ and $ZrB_2$ powders. As an additional control, a PXRD spectrum of the quartz substrate with no polymer or powder film was also collected.

### Density functional theory (DFT) calculations

The calculations were performed using plane-wave basis set within the density functional theory (DFT) with the Vienna Ab-Initio Simulation Package (VASP) package.[85,86] We used the Perdew-Burke-Ernzerhof functional[87,88] with the projector augmented wave (PAW) pseudopotentials.[89,90] We used the DFT-D3 corrections to describe van der Waals interactions.[91] A 500 eV plane-wave cut-off was used, and the systems were relaxed until the forces on the atoms were less than 0.0001 eV/Å. The Brillouin zone was sampled with a Monkhorst-Pack of 15x15x15 k-points in the bulk unit cell, whose equivalent grid was utilized on the layered systems. We used Phonopy[92] to compute the phonon dispersion spectra for the 1L, 2L and bulk $MgB_2$. The electron-energy-loss spectroscopy (EELS) spectra were computed with CASTEP[93] using the optimized geometries from the calculations with VASP, and on on-the-fly generated pseudopotential with PBE functionals. A k-grid of 21x21x1 k-points for all layers and 21x21x21 for the bulk was used in CASTEP. The energy cut-off was 600 eV.

**Supporting Information Available:** Additional sample preparation details, additional TEM images and EELS data, XRD spectra, mechanical data, computational simulation data.

### Author Contributions

MSG, AY, AD, and YG performed the solution phase dispersion experiments. AY and SLYC conducted the HRTEM and EELS measurements and electron scattering simulations. MA and EJGS performed the DFT calculations. MSG, MW, AS, MDG, FT and YL performed the mechanical testing and analysis. XSC, DOL, YG, and SK conducted the AFM imaging. AY, QHW, and AAG wrote the manuscript. QHW, AAG and EJGS directed the project. AAG, AY, and AD have filed a provisional patent related to the results in this manuscript: US15/775,735 and WO2017083693A1 ("Method of preparing metal diboride dispersions and films"), filed November 12, 2015.

**Acknowledgements**

We gratefully acknowledge the use of facilities at the Eyring Materials Center and the W. M. Keck Foundation Laboratory for Environmental Biogeochemistry at ASU, and thank Prof. H. Yan for use of his AFM facilities. AAG, QHW, AY, XSC, and MSG were supported by NSF grant DMR-1610153. AAG and QHW also acknowledge startup funds from Arizona State University (ASU). EJGS acknowledges computational resources through the UK Materials and Molecular Modelling Hub for access to THOMAS supercluster, which is partially funded by EPSRC (EP/P020194/1); and CIRRUS Tier-2 HPC Service (ec019 Cirrus Project) at EPCC (http://www.cirrus.ac.uk) funded by the University of Edinburgh and EPSRC (EP/P020267/1). The Department for the Economy (USI 097) is acknowledged for funding support. MW and MDG acknowledge funding from ARO (grants W911NF-18-1-0412 and W911NF-15-1-0353).